\documentclass[fleqn,usenatbib]{mnras}

\usepackage{newtxtext,newtxmath}

\usepackage[T1]{fontenc}
\usepackage{ae,aecompl}

\usepackage{amsmath}	
\usepackage{amssymb}	
\usepackage{mathtools}
\usepackage{gensymb}
\usepackage{enumitem}

\usepackage{pgf}
\usepackage{soul}
\definecolor{lightgreen}{HTML}{B7F774}
\definecolor{lightred}{HTML}{FF6666}
\definecolor{lightorange}{HTML}{FE9A2E}
\sethlcolor{lightgreen}

\newcommand{\be}{\begin{equation}}
\newcommand{\e}{\end{equation}}
\newcommand{\bear}{\begin{eqnarray}}
\newcommand{\ear}{\end{eqnarray}}
\newcommand{\nline}{\nonumber \\}
\def\dataVec{\mathcal{V}}
\def\responseSet{\mathbfss{T}}
\def\noiseVec{\mathbfit{n}}
\def\regSet{\mathbfss{R}}

\title[Deconvolving the Wedge]{Deconvolving the Wedge: Maximum-Likelihood Power Spectra via Spherical-Wave Visibility Modeling}

\author[A. Ghosh et al.]{
A.~Ghosh,$^{2,3,1}$\thanks{E-mail: aghosh@ska.ac.za}
F.G.~Mertens,$^{1}$
L.V.E~Koopmans$^{1}$
\vspace{0.2cm}\\
$^{1}$Kapteyn Astronomical Institute, University of Groningen, P. O. Box 800, 9700 AV Groningen, The Netherlands\\
$^{2}$Department of Physics and Astronomy, University of the Western Cape, Robert Sobukwe Road, Bellville 7535, South Africa \\
$^{3}$SKA SA, The Park, Park Road, Cape Town 7405, South Africa}

\date{Accepted XXX. Received YYY; in original form ZZZ}

\pubyear{2017}

\begin{document}
\label{firstpage}
\pagerange{\pageref{firstpage}--\pageref{lastpage}}
\maketitle


\begin{abstract}  
Direct detection of the Epoch of Reionization (EoR) via the red-shifted 21-cm
line will have unprecedented implications on the study of structure formation in
the infant Universe. To fulfill this promise, current and future 21-cm
experiments need to detect this weak EoR signal in the presence of foregrounds
that are several orders of magnitude larger. This requires extreme noise control
and improved wide-field high dynamic-range imaging techniques. We propose a new
imaging method based on a maximum likelihood framework which solves for the
interferometric equation directly on the sphere, or equivalently in the
$uvw$-domain. The method uses the one-to-one relation between spherical waves
and spherical harmonics (SpH). It consistently handles signals from the entire
sky, and does not require a $w$-term correction. The spherical-harmonics
coefficients represent the sky-brightness distribution and the visibilities in
the $uvw$-domain, and provide a direct estimate of the spatial power spectrum.
Using these spectrally-smooth SpH coefficients, bright foregrounds can be
removed from the signal, including their side-lobe noise, which is one of the
limiting factors in high dynamics range wide-field imaging. Chromatic effects
causing the so-called ``wedge" are effectively eliminated (i.e. deconvolved) in
the cylindrical ($k_{\perp}, k_{\parallel}$) power spectrum, compared to a power
spectrum computed directly from the images of the foreground visibilities where
the wedge is clearly present. We illustrate our method using simulated LOFAR
observations, finding an excellent reconstruction of the input EoR signal with
minimal bias.
\end{abstract}



\section{Introduction}

The Epoch of Reionization (EoR) is an important milestone in tracing back the
whole history of the Universe. At this epoch the first luminous sources largely
influence the  conditions of the intergalactic medium and therefore played a
significant role in galaxy formation and its evolution. New discoveries of
sources at high redshift (z $\sim$ 8) have pinned down the bright end of the
galaxy luminosity function \citep{Bouwens10, Oesch13}. In parallel, a number of
indirect techniques came up with tight constraints for the redshift of the
ionized to neutral phase transition through cosmic microwave background (CMB)
which tries to constraint reionization from the optical depth of Thomson
scattering to the CMB \citep{PLANCK14}, but the redshift range of the EoR is
significantly less certain. The tail of reionization is also well probed by
Gunn-Peterson absorption troughs in the spectra of high redshift quasars (for a
detailed review see \citet{Fan06}), Ly$\alpha$ emitting galaxies
\citep{Schenker13, Treu13} and the Ly-$\alpha$ absorption profile toward very
distant quasars \citep{Bolton11, Bosman15}. Although, at higher redshifts ($z
\ge 6$) the interpretation of the Lyman-$\alpha$ quasar absorption data becomes
more uncertain. We note that a new population of galaxies discovered by the
various probes still fall well short of reionizing the universe consistently
with the inferred redshifts of the CMB optical depth measurements
\citep{Robertson13, Robertson15}. In addition to these probes, the statistical
technique targeting the 21-cm spin flip transition of neutral hydrogen at high
redshifts has been well recognized as a unique probe of the EoR \citep{Wyithe04,
Bharadwaj05, Furlanetto06, Mesinger10, Morales10, Mellema13} which can reveal
the large-scale fluctuations in the ionization state and temperature of the IGM,
and open up an unique window in the detailed astrophysical processes of the
first sources and their environments.

Recent advances in radio instrumentation and techniques will soon make it
possible to probe the detailed information from the EoR which will enable one to
study structure formation and the formation of the first galaxies. For the
current generation radio telescopes it is believed that a statistical analysis
of the fluctuations in the redshifted 21 cm signal holds significant potential
for observing the HI at high redshifts. Among the foreground sources, discrete
sources can be identified and removed from the images depending on the
sensitivity of the instrument. The contribution from remaining
sources~\citep{DiMatteo02}, the diffuse synchrotron emission from our
Galaxy~\citep{Shaver99} and the free-free emission from ionizing
halos~\citep{Oh03} is still several orders of magnitudes higher compared to the
weak EoR signal.

The foregrounds are expected to have highly-correlated continuum spectra with a
highly correlated spectra. On the contrary, the HI signal is expected to be
uncorrelated at such a frequency separation and there lies the promise of
separating the signal from the foregrounds. In an earlier paper, using a higher
frequency 610 MHz GMRT observation~\citep{Ghosh11b}, we noticed in addition to a
smooth component the measured foreground also had an oscillatory component which
poses a serious problem for foreground removal~\citep{Ghosh11b}. We note that
the angular position of the nulls and the side-lobes of the primary beam (PB)
changes with frequency, and bright continuum sources located near the nulls and
the side-lobes will be seen as oscillations along the frequency axis in the
measured visibilities and subsequently the foreground power spectrum. We showed
in~\citet{Ghosh11b}, one of the possible ways this problem can be reduced by
tapering the array's sky response with a frequency independent window function
that falls off before the first null of the PB pattern and thereby suppresses
the sidelobe response~\citep{Ghosh11b, Choudhuri14}. It is, however,
necessary to note that by tapering the Field of View (FoV) we loose information
at the largest angular scales and secondly, the reduced FoV results in a larger
cosmic variance for the smaller angular modes which are within the tapered FoV.
Moreover, although tapering the FoV is useful in reducing the frequency-
dependent contribution coming from any bright continuum source located near the
nulls of the side-lobes, in general the side-lobe noise arising from varying
point spread function (PSF) of distant bright sources can not be minimized by
just tapering the array's response. The side-lobe noise will always add an extra
noise component to the confusion noise budget of the unresolved sources within
the FoV \citep{Vedantham12}. One of the possible ways to reduce this chromatic
effect is to image a large wide part of the sky and properly account for the
PSF's of the bright far away sources in building up the sky model, and
subtracting it from the visibilities.

Imaging a wide FoV has been tackled traditionally by faceting the sky into a
number of small regions so that we can approximately use tangent planes at the
phase centers of the celestial sphere of each facet to image a wide FoV.
Although, the w-projection algorithm~\citep{Cornwell05, Urbashi09, Bhatnagar13}
has provided sufficient speed improvements over the facet based algorithms,
imaging a very large FoV is non-trivial, specially for aperture arrays which is
sensitive to the entire hemisphere.

It is also important to note that, most of the EoR signal is confined to the
short baselines where the low frequency sky is dominated by Galactic diffuse
emission, confusion noise and side-lobe noise. Therefore, any proper imaging
method has to make a spectrally smooth model of every resolution element of the
sky. Hence, the traditional  ``model-building" of the foregrounds sky is very
inefficient as the sky model is correlated due to the spatially varying PSF. Recently, simulations and analytical calculations have found the existence of a region in cylindrical Fourier space where a part of relatively high $k_{\parallel}$ modes are extensively free of foreground contamination and is well known as ``EoR-window". The boundary of the EoR window is fixed by the intrinsic spectral structure of the foregrounds and the distance between two antennas in wavelengths (baselines). This creates the so called ``foreground wedge" below the EoR window. Basically, the foreground wedge is the effect of increasing mis-alignment of the baselines where the mis-alignment angle is larger for longer baselines \citep{Morales12, Pober13, Dillon15}.
It will be well suited to have a method that models the whole sky or the entire
uvw-volume self-consistently which can be a key step forward. This will also
ensure that side-lobe leakage due to far away sources can be localized well
below the foreground ``wedge" line and thus leaving us with a relatively larger
window to probe the 21-cm EoR signal. We note full-sky interferometric
formulations for aperture arrays has been studied extensively in
\citet{McEwen08} and theoretical ML based formulation has been developed for CMB
\citep{Kim07, Liu16} and recently for transit radio scan telescopes \citep{Shaw14}
where visibilities were represented in spherical harmonic basis. In this paper,
we represent the visibilities in spherical Fourier Bessel basis and used a
Maximum Likelihood (ML) inversion methodology to estimate the corresponding
coefficients in this basis. The simulated foregrounds are modeled as a Gaussian
Random Field (GRF) which has a power spectrum with negative power law index and
assumed to be smooth in frequency. The corresponding visibilities are
represented in spherical Fourier Bessel basis \citep{Carozzi15} where the
coefficients in this basis gives us a direct estimate of the angular power
spectrum in different angular scales. For some smooth foreground template in
frequency, the corresponding spherical Fourier Bessel coefficients is also
expected to have a smooth frequency spectrum. Here, assuming smoothness in
frequency, we implement different foreground removal techniques (such as
polynomial fitting, Principal Component Analysis (PCA), Generalized
Morphological Component Analysis (GMCA) \citep{Chapman13}) which generally tries
to construct a smooth continuum spectra along each line of sight in the
frequency direction and then we use the residual to determine the residual power
spectrum and compare with the input EoR signal, after correction for the noise
bias.

The rest of the paper is organized as follows. In Section 2, we summarize our
methodology and the mathematical formalism. In 3, we describe EoR and 
foreground simulations that has been used in this paper. While in Section 4, we 
elaborate on our foreground cleaning methods and highlights there performance 
with simulated data templates. Finally, in Section 5 we present a summary and 
possible future application of the current work to wide-field effects to next generation upcoming radio interferometers. 

\section{Formalism}

In this section we will introduce a formalism to derive maximum likelihood
estimate of the spherical harmonics coefficients from the sampled visibilities
observed using a radio interferometer.

\subsection{Sky brightness on the celestial sphere and non-coplanar visibilities}

The relationship between the visibility $\mathcal{V}$ and brightness
$B$ on a celestial sphere can be written as~\citep{Thompson01},
\be
\mathcal{V}_{\nu}(\mathbf{r_{\nu}},k) = \int B_{\nu}(\Omega_{k})
\mathrm{e}^{-\mathrm{i}\mathbf{k}\cdot\mathbf{r_{\nu}}} \mathrm{d} \Omega_{k},
\label{eq:VisBright}
\e
where in the visibility domain $\mathbf{r}_{\nu}$ is the separation vector
between two identical receivers, $\mathbf{k}$ is the wave vector and
$\Omega_{k}=(\theta_{k},\phi_{k})$ are the angular components of $\mathbf{k}$ on
the sphere. Here we assume the primary beam of the receivers to be folded into $B_{\nu}(\Omega_{k})$.
Using the Laplace operator, Eqn.~\ref{eq:VisBright} satisfies the three-dimensional
Helmholtz or the spherical wave equation
\be
{\nabla_{r}}^{2}\mathcal{V}+k^{2}\mathcal{V}=0.
\label{eq:LapV}
\e
Along with the Cartesian solutions, the Helmholtz equation has a solution in
spherical coordinates where the eigenfunctions are equal to
\be
j_{\ell}(kr)Y_{\ell m}(\theta,\phi),\quad\textrm{for }\ell=0,1,2,\ldots ,\infty;\: m=-\ell,\ldots,\ell.
\label{eq:jlYlsph}
\e
Here, $Y_{\ell m}(\Omega)$ is the standard orthonormal spherical harmonic
function and $j_{\ell}(kr)$ is the spherical Bessel function of the first kind.
Following~\citet{Carozzi15}, Eqn.~\ref{eq:VisBright} can be recast into the
eigen function of Eqn.~\ref{eq:jlYlsph}. Using the Legendre addition theorem and
the Jacobi-Anger expansion of a plane wave, one obtains,
\be
\mathrm{e}^{-\mathrm{i}\mathbf{k}\cdot\mathbf{r}}=4\pi\sum_{lm}(-\mathrm{i})^{\ell}j_{\ell}(kr)Y_{\ell m}(\theta_{r},\phi_{r})Y_{\ell m}^{\ast}(\theta_{k},\phi_{k}),
\label{eq:PlaneWave}
\e
Here, and subsequently, we use the short-hand notation $\sum_{lm} = \sum_
{\ell=0}^{\infty}\sum_{m=-\ell}^{\ell}$.
Using this relation in Eqn.~\ref{eq:VisBright} we find,
\be
\mathcal{V} =\displaystyle \int B(\Omega_{k})\left(4\pi\sum_{lm}(-\mathrm{i})^
{\ell}j_{\ell}(kr)Y_{\ell m}(\theta_{r},\phi_{r})Y_{\ell m}^{\ast}(\Omega_{k})\right)\mathrm{d}\Omega_{k}.
\label{eq:vCZplanewave}
\e
Similar to the visibilities, the sky brightness distribution over a celestial
sphere can be expanded in the spherical harmonic basis
\be
B(\Omega_{k})=\sum_{lm}b_{\ell m}Y_{\ell m}(\Omega_{k}),
\label{eq:ImgYlm}
\e
where $b_{lm}$ are the multipole moments of the sky. Using Eqn. \ref{eq:vCZplanewave} we find,
\bear
\mathcal{V}&=&4\pi\sum_{lm}(-\mathrm{i})^{\ell}j_{\ell}(kr)Y_{\ell m}(\theta_{r},\phi_{r})\, \nline
&& \times\int\left(\sum_{lm}b_{\ell m}Y_{\ell m}(\Omega_{k})\right)Y_{\ell m}^{\ast}(\Omega_{k})\mathrm{d}\Omega_{k}\, \nline
&=&4\pi\sum_{lm}(-\mathrm{i})^{\ell}j_{\ell}(kr)Y_{\ell m}(\theta_{r},\phi_{r})b_{\ell m},
\label{eq:cartVisEqSphBri}
\ear
where we use the orthogonality relation for the spherical harmonic functions
\be
\int_{0}^{4\pi}Y_{\ell m}(\Omega)Y_{\ell'm'}^{\ast}(\Omega)\mathrm{d}\Omega=\delta_{\ell\ell'}\delta_{mm'}.
\e
It follows from Eqn.~\ref{eq:LapV} that the visibility distribution in spherical
coordinate can be expanded in terms of the eigenfunction of the spherical wave
equation with coefficients $\tilde{v}_{\ell m}$:
\be
\mathcal{V}=\sum_{lm}\tilde{v}_{\ell m}j_{\ell}(kr)Y_{\ell m}(\Omega_{k}).
\label{eq:VisibilitySphDecomp}
\e
Now comparing Eqn.~\ref{eq:cartVisEqSphBri} and
\ref{eq:VisibilitySphDecomp} and using the orthonormality relation of
the $Y_{\ell m}$ harmonics, we find the $(\ell,m)$ coefficients in the
sky and the visibility domain are one to one related by \citep{Carozzi15},
\be
\tilde{v}_{\ell m}=4\pi(-\mathrm{i})^{\ell}b_{\ell m}.
\label{eq:vlmblm}
\e
This shows that there is a simple proportionality relation between the
brightness distribution, in terms of $b_{\ell m}$, and the visibility
distribution, in terms of $\tilde{v}_{\ell m}$. While the former is defined on a
2D sphere, the latter is related to the 3D uvw-domain, but as in holography,
these 2D and 3D spaces contain identical information. We also note that the
spherical harmonic components are eigenfunctions of the measurement
equation on the sphere and the components satisfy the Helmholtz
dispersion relation $k^{2}=\omega^{2}/c^{2}$. On the other hand, plane
wave solution of the Cartesian Fourier transform are not
eigenfunctions of the measurement equation on the sphere and does not
satisfy the dispersion relation leading to the additional complexity
of dealing with the $w$-term of the wave-vector.

\subsection{Spherical harmonics of a real sky}

Next, extending ~\cite{Carozzi15}, we investigate whether we can
simplify Eqn. \ref{eq:VisibilitySphDecomp} for a real sky. We note that
the positive and negative $m$ modes for a real sky are related by,
\be
{b}_{\ell -m} = (-1)^{m} {b}_{\ell m} ^{*},
\label{eq:blm}
\e
and we also have the orthogonality relation between the spherical harmonics as,
\be
{Y}_{\ell -m}(\Omega_{k}) = (-1)^{m} {Y}_{\ell m} ^{*}(\Omega_{k}).
\label{eq:ylm}
\e
Combining Eqn. \ref{eq:vlmblm}, \ref{eq:blm} and \ref{eq:ylm} we find,
\be
\tilde{v}_{\ell -m} = (-1)^{m} (-1)^{l}  \tilde{v}_{\ell m}^{*}.
\label{eq:vlm}
\e
From Eqn. \ref{eq:vlm} we notice that $ \tilde{v}_{\ell 0} = (-1)^{l}
\tilde{v}_{\ell 0}^{*} $. This signifies that for $m = 0$ mode
$\tilde{v}_{\ell 0}$ modes are real for even $\ell$ and for odd $\ell$ it is imaginary.
Using further algebraic manipulations we can show that,
\begin{align}
\tilde{v}_{\ell m}{Y}_{\ell m}(\Omega_{k}) +&  \tilde{v}_{\ell -m}{Y}_{\ell -m}
(\Omega_{k})
= \tilde{v}_{\ell m}{Y}_{\ell m}(\Omega_{k}) + (-1)^{\ell}
\tilde{v}_{\ell m}^{*}{Y}_{\ell m}^{*}(\Omega_{k}) \, \nline
=&
\begin{dcases}
2\,\Re(\tilde{v}_{\ell m}{Y}_{\ell m}(\Omega_{k})) &  \rm{for \; \ell \;
even}\,, \\
2 \mathrm{i}\,\Im(\tilde{v}_{\ell m}{Y}_{\ell m}(\Omega_{k}))  & 
\rm{for \; \ell \; odd}. \\
\end{dcases}
\label{eq:vlmylm}
\end{align}
Folding this relation into the visibility equation \ref{eq:VisibilitySphDecomp}
implies that the real part of the visibilities are composed of even $\ell$ modes
and the imaginary part are composed of odd $\ell$ modes.

\begin{equation}
\begin{split}
\mathcal{V} & = \sum_{\ell=2p} j_{\ell}(kr) \big[\Re(\tilde{v}_{\ell 0}Y_{\ell
0}(\Omega_{k})) \\  &\qquad\qquad + 2 \sum_{m>0} \Re(\tilde{v}_{\ell m}Y_{\ell
m}(\Omega_{k})) - \Im(\tilde{v}_ {\ell m}Y_{\ell m} (\Omega_{k})) \big] \\
&\quad +\mathrm{i}\sum_{\ell=2p+1} j_{\ell}(kr) \big[\Im(\tilde{v}_{\ell
0})\Re(Y_{\ell 0}) (\Omega_{k})) \\ &\qquad\qquad + 2 \sum_{m>0}
\Im(\tilde{v}_{\ell m})\Re(Y_{\ell m}(\Omega_ {r})) - \Re(\tilde{v}_ {\ell
m})\Im(Y_{\ell m} (\Omega_{k})) \big] 
\label{eq:Visibility_levlodd}
\end{split}
\end{equation}

An interesting consequence of this equation is that even and odd $\ell$ modes
can be recovered independently from the real and imaginary part of the
visibilities. This effectively reduces the computation time for inverting Eqn.
\ref{eq:VisibilitySphDecomp} and subsequently determining the $\tilde{v}_{\ell
m}$ coefficients.


\subsection{Maximum-Likelihood Inversion}
\label{sec:ml_inversion}

In this section, we present our ML solutions of $\mathbfit{v}_{\mathrm{ML}}$
based on the visibility data that we generated. We note that \citet{Kim07} have
studied the direct reconstruction of spherical harmonics from visibilities for
cosmic microwave background analysis where they compute the maximum likelihood
solutions for the SpH coefficients directly from visibilities without going into
the map space. Their analysis is mainly restricted to two dimensions whereas our
analysis also incorporate the line of sight `w' component of each visibility.
Here, we represent the visibility data in the form of a system of linear
equations,
\be
\dataVec =  \responseSet \mathbfit{v}_{\mathrm{ML}} + \noiseVec ,
\label{eq:VisibMat}
\e
where $\responseSet$ is the transformation matrix which includes the spherical
Bessel basis function, $\noiseVec$ is the Gaussian random noise in each
visibility with mean zero and covariance $\left\langle \noiseVec
\noiseVec^{\mathrm{H}} \right\rangle \equiv \mathbfss{C}_{\mathrm{D}}$. This
equation is a simple translation of Eqn.\ref{eq:Visibility_levlodd} in matrix
form and hence  Eqn.\ref{eq:Visibility_levlodd} is used to build the
transformation matrix $\responseSet$. In principle the full sky is described by
an infinite number of $(\ell, m)$ modes and hence a proper sampling needs to be
chosen which will be discussed in section~\ref{sec:sampling}.

In this paper, we have used a conjugate gradient (or quasi-Newtonian) optimization
scheme to solve the minimum variance estimator of Eqn.~\ref{eq:VisibMat}, which is given by
\citep{Tegmark97},
\be
\label{minvarest}
\mathbfit{v}_{\mathrm{ML}} = (\responseSet^{\mathrm{H}}\mathbfss{C}_{\mathrm{D}}^{-1} \responseSet)^{-1} \responseSet^{\mathrm{H}} \mathbfss{C}_{\mathrm{D}}^{-1} \dataVec,
\e
with a error co-variance matrix for $\mathbfit{v}$,
\be
\mathbf{\Sigma_{\mathbfit{v}}} = (\responseSet^{\mathrm{H}} \mathbfss{C}_{\mathrm{D}}^{-1}
\responseSet)^{-1}.
\e
We note in many cases for interferometric data sets, the problem of finding the
most likely (ML) solutions is ill-posed and we need to introduce priors to
regularize the solution of $\mathbfit{v}_{\mathrm{ML}}$. With regularization the
new form of the ML solutions and the error co-variance matrix get updated as
\citep{Ghosh15,Zheng17},
\be
\mathbfit{v}_{\mathrm{ML}}^{\regSet} = (\responseSet^{\mathrm{H}}\mathbfss{C}_{\mathrm{D}}^{-1} \responseSet + \regSet)^{-1} \responseSet^{\mathrm{H}} \mathbfss{C}_{\mathrm{D}}^{-1} \dataVec,
\e
where, $\regSet$ is the regularization matrix. With the introduction of
$\regSet$ the new error co-variance matrix is,
\be
\mathbf{\Sigma}^{\regSet} = (\responseSet^{\mathrm{H}}\mathbfss{C}_{\mathrm{D}}^{-1} \responseSet + \regSet)^{-1}(\responseSet^{\mathrm{H}}\mathbfss{C}_{\mathrm{D}}^{-1} \responseSet)(\responseSet^{\mathrm{H}}\mathbfss{C}_{\mathrm{D}}^{-1} \responseSet + \regSet)^{-1},
\label{eqerrorcov}
\e
where $(\responseSet^{\mathrm{H}}\mathbfss{C}_{\mathrm{D}}^{-1} \responseSet + \regSet)^{-1}(\responseSet^{\mathrm{H}}\mathbfss{C}_{\mathrm{D}}^{-1} \responseSet)$ acts as a point spread function for the corresponding $(\ell,m)$ mode in the true sky map \citep{Zheng17}.

In general the computational effort of these linear inversion problem is in
order $\sim \rm{N}_{lm}^3$, where $\rm{N}_{lm}$ is the number of modes in the
sky that we are interested in. This leads to a substantial floating point
operations per frequency channel. To overcome it, we run our ML inversion on a
196-CPU parallel 2TB shared memory machine. We note that currently creating the
transformation matrices from the baseline co-ordinates takes more time than
solving the system of linear equations, because of the computationally expensive
spherical-harmonic and Bessel functions and the large size of the matrices
involved.

\subsection{Sampling the spherical harmonics}
\label{sec:sampling}

In general 21-cm power-spectra analyses are only done on short interferometric
baselines, where its signal to noise is expected to be largest. This allows us
to limit the range of $(\ell,m)$ modes that constitute the matrix $\mathrm{T}$. As
illustration, for the LOFAR-EoR project, we use the LOFAR baseline range between
50$\lambda$ - 250$\lambda$ \citep{Patil17}. This translates into recovering spherical
harmonics modes from $\ell_{\mathrm{max}} = 314$ to $\ell_{\mathrm{max}} =
1570$, which implies a total of 1\,185\,351 modes to recover the full sky.
Such a large inversion would still be intractable, but we can further reduce the
number of coefficients to estimate, considering that we are observing the sky
modulated by the primary beam (PB) of the telescope.
The observed sky brightness $B(\Omega_k)$ can be expanded as
\be
B(\Omega_k) = H(\Omega_k) A(\Omega_k),
\e
with $A(\Omega_k)$ is the full-sky brightness and $H(\Omega_k)$ being the PB. We can
always set the phase center, and hence the center of the PB, to be at $\Omega_k
= (0\,, 0)$. Assuming an axi-symmetrical PB\footnote{This is only valid at
a first order, especially with phased-array telescopes. Nevertheless, this
approximation is good enough for our purpose of deriving a spherical harmonics
sampling rule.}, this simplify the PB function to be a function of $\theta_k$
only, $H(\Omega_k) = H(\theta_k)$.
The spherical harmonics basis can be decomposed as
\be
 Y_{\ell m}(\theta_k, \phi_k) = N_{\ell m} P^m_{\ell}(\mathrm{cos}(\theta_k))
 \mathrm{e}^{\mathrm{i} m \phi},
\e
with $N_{\ell m}$ a normalization factor and $P^m_{\ell}$ the associated
Legendre functions, as function of the cosine angle. Expanding the spherical
harmonics coefficients of the observed sky brightness $b_{\ell m}$ into this
basis, we obtain
\be
b_{\ell m} = \int_{0}^{\pi}\left[\int_{0}^{2 \pi} A(\Omega_k) \mathrm{e}^{
\mathrm{i} m \phi} d \phi_k \right] H(\theta_k) N_{\ell m} P^m_{\ell}(
\mathrm{cos} (\theta_k)) d \theta_k,
\label{eq:sph_transform}
\e
\begin{figure}
    \includegraphics{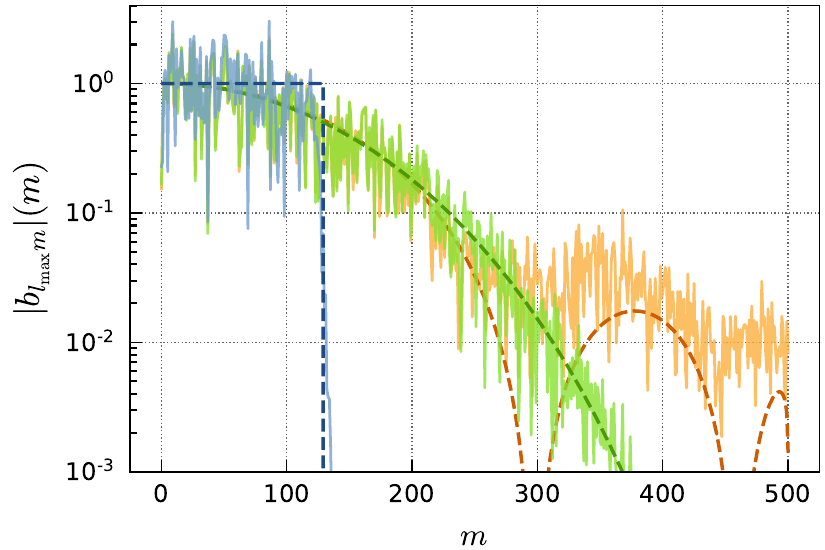}
    \caption{\label{fig:blm_beam_effect_m} 
    Effect of a primary beam on the $b_{\ell m}$. The norm of the beam modulated
spherical harmonics coefficients    of a
    simulated flat sky (solid line) is plotted for a fixed
    $\ell$ mode ($\ell = \ell_{
    \mathrm{max}}$) and compared against the profile of the primary beam
     $H(\mathrm{sin}^{-1}(\frac{m}{{\ell}}))$ (dashed line) for a
    top-hat (blue), Gaussian (green) and Bessel (orange) primary beam.}
\end{figure}
\begin{figure}
    \includegraphics{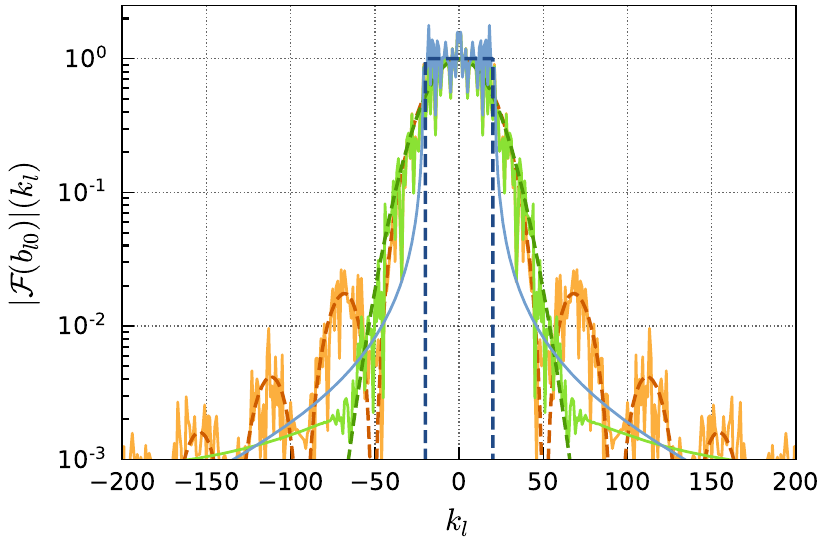}
    \caption{\label{fig:blm_beam_effect_l}
    Effect of a primary beam on the $F_{\ell}[b_{\ell m}]$. The norm of $F_
    {\ell}[b_{\ell m}]$ of a simulated flat sky (solid line) is plotted for a
    fixed $m$ mode ($m = 0$) and compared against the profile of the
    primary beam $H(\frac{2
    k_{\ell}}{\ell_{\mathrm{max}} - m} \pi)$ (dashed line) for a top-hat (blue), 
    Gaussian (green) and Bessel (orange) primary beam.}
\end{figure}
which means that modulating the sky brightness by an axi-symmetric primary beam
$H(\Omega_k)$ is equivalent to modulating the associated Legendre functions.
Exploring this relation, we find that multiplying the sky by a beam can be
viewed as a convolution in $\ell$, and a multiplication in $m$. More
specifically, for a fixed $\ell$, we have
\bear
\label{eq:blm_pb_relation_m}
b_{\ell m} &\propto& a_{\ell m} H(\mathrm{sin}^{-1}(\frac{m}{{\ell}})), \\
\label{eq:blm_pb_relation_l}
\mathcal{F_{\ell}}[b_{\ell m}](k_{\ell}) &\propto& \mathcal{F_{\ell}}[a_{\ell
m}](k_{\ell}) H(\frac{2 \pi k_{\ell}}{\ell_{\mathrm{max}} - m}),
\ear
\begin{figure*}
    \includegraphics{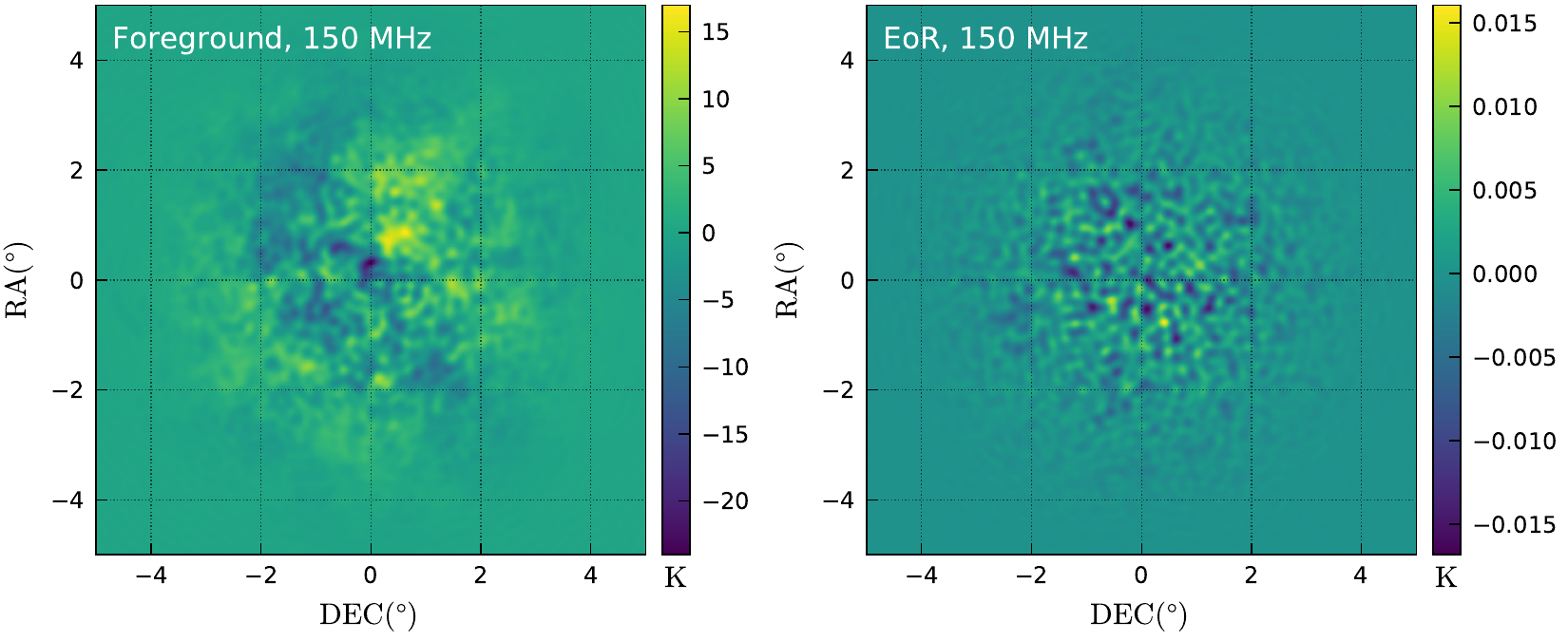}
    \caption{\label{fig:map_fg_eor_templates_150MHz} This figure shows a slice at 150 MHz for the diffuse foreground emission and the 21-cm eor signal template at a pixel resolution of 1.17 arc-min. Note that, the mean is subtracted from the maps. Each panel covers a region of $5^{\circ} \times 5^{\circ}$.}
\end{figure*}
with $\mathcal{F_{\ell}}$ denoting the Fourier transform in $\ell$ (for a fixed
$m$), $k_\ell$ the Fourier conjugate of $\ell$, $b_{\ell m}$ the
spherical harmonics coefficient of the beam modulated
sky, and $a_{\ell m}$ the spherical harmonics coefficient of the sky.

To experimentally confirm these relations, we simulate a Gaussian Random Field
(GRF) on the sphere with a flat power spectra ($C_{\ell} = 1$ for all $\ell$).
This simulated sky is multiplied by a primary beam and we then compute the
associated spherical harmonics coefficient using Eqn.~\ref{eq:sph_transform}.
For this test, a top-hat,  a Gaussian and a Bessel primary beam are used. In
Fig.~ \ref{fig:blm_beam_effect_m}, the $b_{\ell m}$ (solid lines) are compared
with the profile of the primary beam $H(\mathrm{sin}^{-1}(\frac{m}{{\ell}}))$
(dashed line), and in Fig.~\ref{fig:blm_beam_effect_l}, the
$\mathcal{F_{\ell}}[b_{\ell m}](k_{\ell})$ are compared with $H(\frac{2 \pi k_
{\ell}}{{\ell} - m})$. For the three beams, these experiments confirm that
Eqn.~\ref{eq:blm_pb_relation_m} and Eqn.~\ref{eq:blm_pb_relation_l} are good
approximation of the relation between the $b_{\ell m}$ and the primary beam
profile $H(\theta_k)$.

While we could not obtain an exact relation, these approximations are sufficient
to define simple sampling rules. Reducing the sky Field of View (FoV)
corresponds in the spherical-harmonics domain (defined at the phase-center) to
reducing $m_ {\mathrm{max}}(\ell)$, the maximum $m$ mode as function of $\ell$,
and increasing $\Delta \ell$, the spacing between
consecutive $\ell$ modes. Eqn.~\ref{eq:blm_pb_relation_l} suggests that these
sampling rules are better applied in Fourier domain in $\ell$. Updating
Eqn.~\ref{eq:VisibMat} accordingly, we have,
\bear
\label{eq:VisibMatFT}
\dataVec &=&  \responseSet \mathbfss{F}_{\ell} \mathbfit{V}_{\mathrm{ML}} + \noiseVec, \\
\label{eq:VlmFT}
\mathbfit{v}_{\mathrm{ML}} &=&  \mathbfss{F}_{\ell}^{-1} \mathbfit{V}_{
\mathrm{ML}},
\ear
with $\mathbfss{F}$ denoting the matrix form of the Fourier transform in $\ell$.
We now solve for the $\mathbfit{V}_{\mathrm{ML}}$ using eqn.~\ref{eq:VisibMatFT}
and then compute the $\mathbfit{v}_{\mathrm{ML}}$ using eqn.~\ref{eq:VlmFT}. The
size of $\mathbfit{V}_{\mathrm{ML}}$ can be reduced so that it contains only
significant elements. Using Eqn.~ \ref{eq:blm_pb_relation_m} and
Eqn.~\ref{eq:blm_pb_relation_l}, we can demonstrate that to fully describe the
beam modulated sky up to $\theta_{
\mathrm{max}}$, we can restrict the $k_{\ell}(m)$ modes of $\mathbfit{V}_{
\mathrm{ML}}$ such that
\bear
|k_{\ell}(m)| &<=& \frac{\theta_{\mathrm{max}} (\ell_{\mathrm{max}} - m)}{2 \pi},
\\
m(l) &<=& \ell \, \mathrm{sin}(\theta_{\mathrm{max}}).
\ear

Using these sampling rules considerably reduce the computational scale of the
problem. If we take for example a Gaussian primary beam with full width at half
maximum (FWHM) $\theta_{\mathrm{fwmh}}$, it is reasonable to solve only for the
($\ell$, m) coefficients restricted to $\theta_{\mathrm{max}} \sim 2 \theta_{
\mathrm{fwmh}}$ without impacting the inversion strongly. For a FWHM of $4
\degree$ and $\ell_{\mathrm{max}} = 1570$, this reduce the number of
coefficients to solve for from 1\,185\,351 to 7\,864.

\section{Full simulation}

\subsection{Simulated data templates}

In this section we describe briefly the templates we considered for the EoR
signal and diffuse foreground emission from which the visibility data were
generated. In our formalism we assume that the bright extragalactic sources can
be properly modeled and subtracted from the data, so they are not included in
our foreground model.

\subsubsection{EoR Signal}

We have used the semi-analytic code 21cmFAST \citep{Mesinger07,Mesinger11} to
simulate the EoR signal. 21cmFAST treats physical processes with approximate
methods. Apart from the scales less than $< 1 {\rm Mpc}$, the output of this
semi-analytic code  tends to agree well with the hydro-dynamical simulations of
\citet{Mesinger11}. The 21-cm EoR template used here is the same as used in
\citet{Ghosh15} and we refer the reader to \citet{Chapman12} for details
description of the simulations.
The 21cmFAST simulation computes the $\delta T_{\rmn{b}}$ box at each redshift
based on the following equation,
\begin{eqnarray}
\begin{split}
 \delta T_\mathrm{b} &= 28\: \mathrm{mK} \times (1+\delta)x_{\mathrm{HI}}\left(1-\frac{T_{\mathrm{CMB}}}{T_{\mathrm{spin}}}\right)\left(\frac{\Omega_b h^2}{0.0223}\right) \\ &\qquad \times \sqrt{\left(\frac{1+z}{10}\right)\left(\frac{0.24}{\Omega_m}\right)},
\end{split}
\end{eqnarray}
where, $\delta T_\mathrm{b}$ is the brightness temperature fluctuation which is
detected as a difference from the background CMB temperature $T_{\mathrm{CMB}}$
\citep{Field58,Field59,Ciardi03}, $h$ is the Hubble constant in units of $100\:
\mathrm{km s}^{-1}\:\mathrm{Mpc}^{-1}$, $x_{\mathrm{HI}}$ is the neutral
hydrogen fraction and $\Omega_b$ and $\Omega_m$ are the baryon and matter
densities in critical density units. We note that here we ignore the gradient of
the peculiar velocity fluctuation whose contribution to the brightness
temperature is relatively small \citep{Ghara14, Shimabukuro15}. We also assume
that the  neutral gas has been heated well above the CMB temperature during
Epoch of Reionization ($T_{\rmn{S}} \gg T_{\rmn{CMB}}$) \citep{Pritchard08} and
therefore we can safely neglect the spin temperature fluctuations in generating
the simulated 21-cm signal.

\subsubsection{Diffuse foregrounds}

The diffuse foreground model used in this paper include contribution from
Galactic diffuse Synchrotron emission (GDSE), Galactic localized Synchrotron
emission, Galactic diffuse free-free emission and unresolved extra-galactic
foregrounds. We refer the reader to \citet{Jelic08,Jelic10} for a  detail
comprehensive review how the individual foreground components were simulated.
Here, we highlight only few key features of the foreground simulations.
Galactic diffuse Synchrotron emission (GDSE) originates due to the interaction
of cosmic ray (CR) electrons produced mostly by supernova explosions and the
Galactic magnetic field \citep{Pacholczyk70, Rybicki86}. The intensity and the
spectral  index of the GDSE are modeled as Gaussian random fields (GRF) where
the spatial power spectrum had a power law index  of $-2.7$ and with frequency
the GRF assumes a spectral index of $-2.55 \pm 0.1$ \citep{Shaver99} with a
fixed mean brightness temperature set around to $253 \pm 1.3$ K at 120 MHz.
The diffuse thermal (free-free) emission arises due to bremsstrahlung radiation
in very diffuse ionized gas. At LOFAR-EoR frequencies the ionized gas is
optically thin and the free-free emission from diffuse ionized gas is
proportional  to the emission measure. Here, the free-free emission is modeled
as a GRF and spectral index is fixed to -2.15 \citep{Tegmark00,Santos05} where
the normalization is set with respect to the intensity of the $\rm{H}\alpha$
emission and fixed at $2.2$ K at 120 MHz \citep{Smoot98}.
The simulations of radio galaxies, used in this paper, are
based on the extragalactic radio source counts at 151 MHz by \citet{Jackson05}.
Then the simulated radio galaxies are clustered using a random walk algorithm where the radio clusters are selected from the cluster catalogue of Virgo
Consortium\footnote{http://www.mpa-garching.mpg.de/galform/virgo/hubble/}.

This diffuse foreground model is finally calibrated to have a spatial power
spectra of 400 $\rm{mK}^2$ at 150 MHz and $\ell = 400$, to match more closely
with LOFAR \citep{Patil17} and Westerbork \citep{Bernardi10} observations of the
diffuse emission on the NCP field.

\begin{figure*}
    \includegraphics{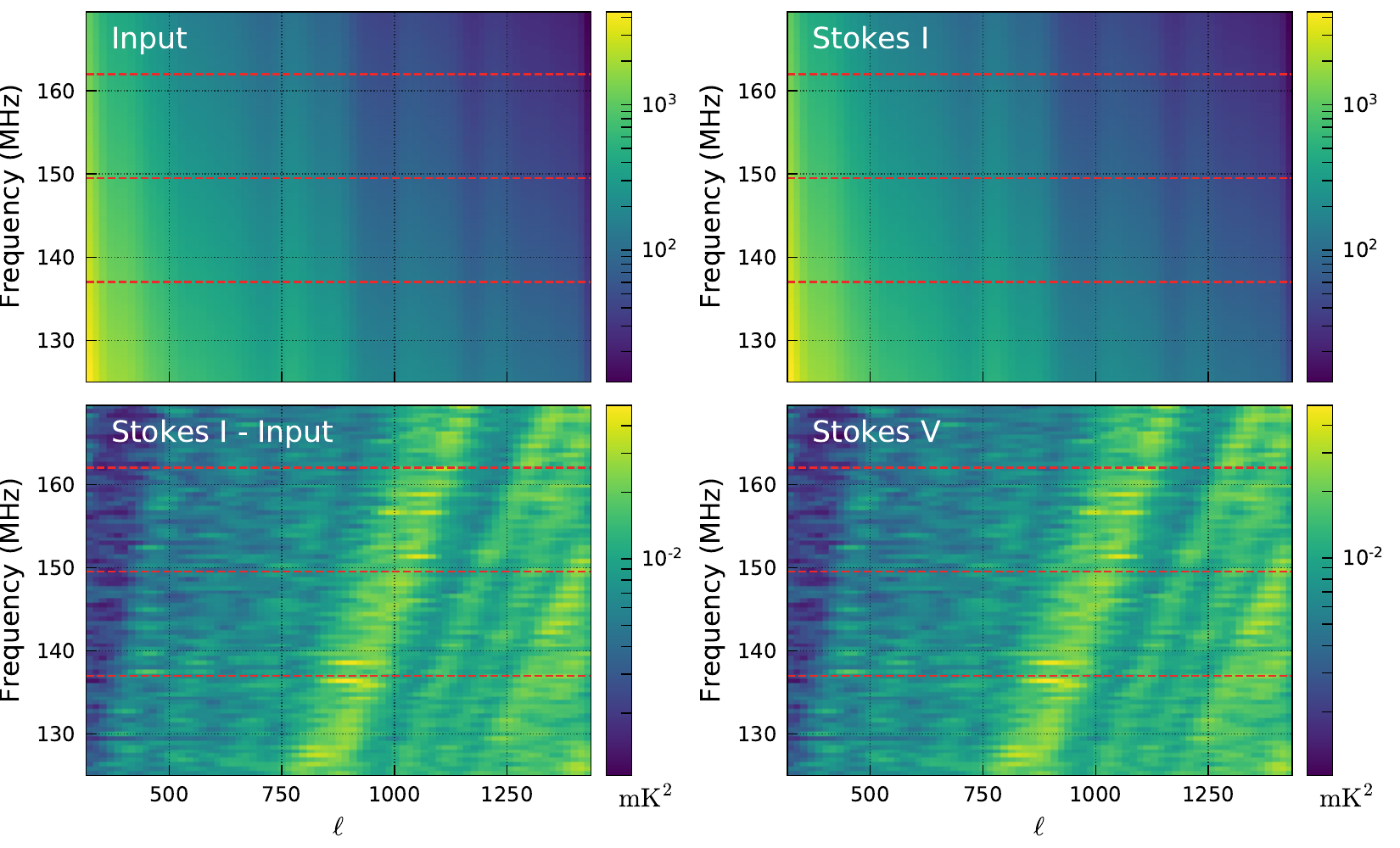}
    \caption{\label{fig:c_l_inprec} This figure shows the angular power
    spectrum as a function of angular scale $\ell$ and frequency $\nu$ for the
    input simulated sky composed of foregrounds diffuse emission and 21-cm
    signal (top-left panel), the Spherical-Harmonics(SpH) ML reconstructed  
    Stokes I (top-right panel), the difference between the input sky and Stokes I 
    (bottom-left) and SpH ML reconstructed Stokes V (bottom-right panel). The Stokes
    V power is due to the noise in the visibilities, and is similar to the difference 
    in power between Stokes I and Input map. The red dashed line delimit the two 
    frequency bins, 137.5-150 MHz ($z \sim 9$) and 150-162.5 MHz ($z \sim 8$) at 
    which the three-dimensional power spectra are computed.}
\end{figure*}

\subsection{Simulating visibilities}

The LOFAR-HBA antenna positions \citep{vanHaarlem13} were used to generate the
baseline components $(u, v, w)$ towards the North Celestial Pole (NCP) at which
we  predict visibilities corresponding to a combination of diffuse foregrounds
and EoR only sky. Figure \ref{fig:map_fg_eor_templates_150MHz} shows a
representative spatial slice at 150 MHz of the foreground and the EoR signal
used in this simulation. We used a Gaussian primary beam with a FWHM of $4
\degree$ \citep{vanHaarlem13} to model the LOFAR primary beam which is
multiplied with the input foreground and the 21-cm signal templates of $10
\degree \times 10 \degree$. The Cartesian maps were converted to spherical
harmonics using the \textsc{healpix} \footnote{http://healpix.jpl.nasa.gov}
package, and then transformed to visibilities using
Eqn.~\ref{eq:VisibilitySphDecomp}. Next, we added random Gaussian noise to the
real and imaginary part of the visibility separately where the rms of the noise
were calculated from,

\be 
\sigma = \frac{\rm{SEFD}}{\sqrt{2 \, \Delta \nu \, \Delta t}},
\e 

with $\Delta \nu$ and $\Delta t$ the frequency bandwidth and integration time,
respectively. We assume that for LOFAR HBA the expected system equivalent flux
density (SEFD) towards NCP is $\sim 4000$ Jy~\citep{vanHaarlem13}. We note that
the SEFD is generally elevation dependent and changes across the sky. We added
Gaussian random noise with rms of 0.04 Jy, which corresponds to about 100 nights
of 12 hours long LOFAR observation for a 0.5 MHz channel width and 100 s
snapshot integration time. This integration time is chosen to reduce the number
of visibilities and hence the complexity of the ML inversion, while avoiding
time-smearing effect for the selected baseline range.

We simulate both visibilities including the sum of the foregrounds and the
21-cm signal input data template, which will be our Stokes I data set, and the
visibilities of the noise only, which will be our Stokes V data set.

\subsection{ML inversion and Power Spectra}

From the simulated Stokes-I and Stokes-V visibilities, we infer the
recovered spherical harmonics $b^{\mathrm{I}}_{\ell m}$ and
$b^{\mathrm{V}}_{\ell m}$, using the maximum likelihood algorithm described in
Sect.~\ref{sec:ml_inversion} and Sect.~\ref{sec:sampling} and compute the
angular power spectra via,
\be
C_{\ell} = \frac{4 \pi}{\Omega_{\mathrm{PB}}} \frac{1}{\ell + 1} \sum_{m = 0}^
{\ell} |b_{\ell m}|^2,
\e
with $\Omega_{\mathrm{PB}}$ being the primary beam field of
view~\citep{Aaron12}. Figure~ \ref{fig:c_l_inprec} shows the angular power
spectra as function of $\ell$ and frequency for the input sky and for the
recovered Stokes I and Stokes V. We find the input and the reconstructed angular
power spectra for all the frequency channels resembles each other quite closely
and this estimator can be used to jointly characterize the angular and frequency
dependence of the observed sky signal. It also shows that the error introduced
by the ML inversion is well below the thermal noise, as the power of the
difference between the Stokes I and input sky is found to be similar to the
Stokes V power. The diagonal structures observed in the Stokes V power
spectra are related to baseline density. The noise is higher for $\ell$ modes
corresponding to sparser baseline density which is a function of fixed baseline
metric units and then scale to units of lambda as a function of frequency: $\ell
= \frac{2 \pi |\mathbfit{b}|}{\lambda}$ where \mathbfit{b} is the vector
representing the coordinates in meters in the plane of the array.


 \begin{figure}
    \includegraphics{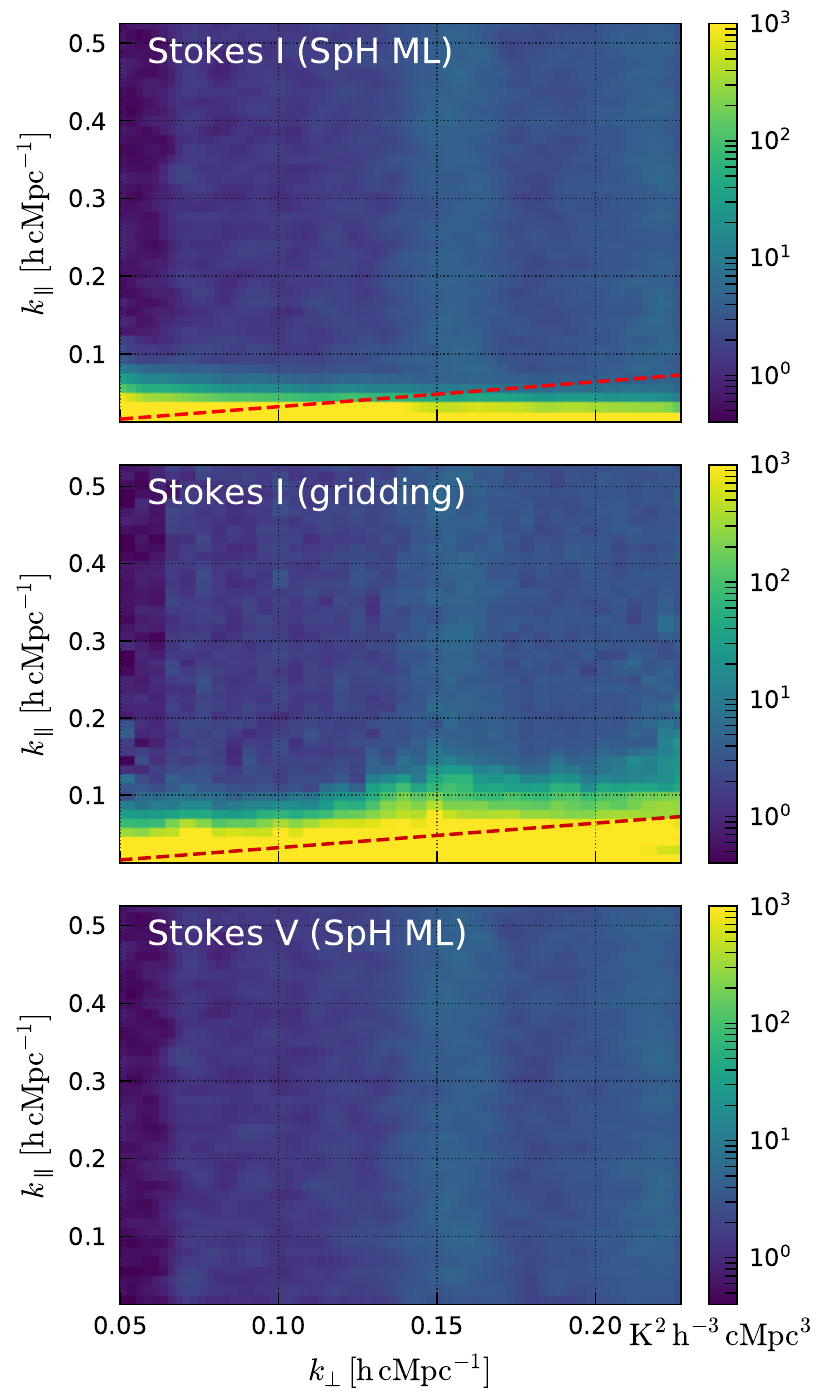}
    \caption{\label{fig:ps2d_I+V_80sb}
Cylindrically averaged power spectra estimated from the simulated visibilities.
The top and middle panel shows the power spectra of the observed signal
(foregrounds, noise and 21 cm signal; Stokes I) obtained from the spherical
harmonics ML inversion (top) and by gridding the simulated visibilities using
WSClean (middle). The absence of structure inside the 10 $\degree$ Field of View
wedge line (red dashed line) in the power spectra estimated using the SpH ML
inversion demonstrate that the method effectively compute PSF-deconvolved
representation of the sky. The power spectra of the noise (Stokes V) is plotted
in the bottom panel.
    }
 \end{figure}

We then introduce the three-dimensional power spectra to quantify the entire
second order statistics of the background sky signal by taking the Fourier
transform of the $b_{\ell m}(\nu)$ cube in the frequency direction.
We define the cylindrically averaged power spectra as~\citep{Aaron12}:
\begin{equation}
P(k_{\perp}, k_{\parallel}) = \frac{4 \pi X^2 Y}{\Omega_{\mathrm{PB}} B} 
\frac{1}{\ell + 1} \sum_{m = 0}^{\ell} \left|\hat{b}_{\ell m} (\eta)\right|^2,
\end{equation}
where $\eta$ is the Fourier conjugate of $\nu$, $\hat{b}_{\ell m} (\eta)$ the Fourier transform of the cube $b_{\ell m}
(\nu)$, B the frequency bandwidth, X and Y are conversion factors from angle and
frequency to comoving distance, and where the Fourier modes are in units of
inverse comoving distance and are given by~\citep{Morales06,Trott12}:
\begin{align}
k_{\perp} &= \frac{\ell}{D_M(z)},\\ k_{\parallel} &= \frac{2 \pi H_0 f_{21} E
(z)}{c(1+z)^2} \eta,
\end{align}
with $D_M(z)$ the transverse co-moving distance, $H_0$ the Hubble constant, $f_
{21}$ the frequency of the hyperfine transition, and $E(z)$ the dimensionless
Hubble parameter~\citep{Hogg10}.

It is important to point out that the three-dimensional power spectrum $P(k)$ is
well suited to quantify the statistics of HI signal only if we perform it on
limited frequency ranges which prevent the evolution of the HI signal across the
line of sight, known as the Light-Cone effect \citep{Kanan12}, assuming the
21-cm signal to be statistically stationary in angular and frequency axis
\citep{Mondal17,viswesh17}. For this reason, we limit our estimation of the
three dimensional power spectrum to frequency bands of 12.5 MHz. For the
foregrounds, where the statistics are quite different for the chromatic response
of the telescope the angular and frequency homogeneity assumption breaks down
and the power spectrum $P(k)$ is no longer the obvious choice to quantify the
statistical properties of the measured sky signal. For this latter case, the
angular power spectrum estimator ($C_{\ell}$'s) as a function of frequency is
more suitable. We note that this assumes homogeneity in angular domain but does
not rely on the assumption of homogeneity in the frequency domain. We also note
that when using the visibility correlation \citep{Bharadwaj01, Bharadwaj05,
Ghosh11b, Ali08, Ghosh12}, the angular power spectrum has been quantified
earlier and the relation between the visibility correlations and the power
spectrum $P({\bf {\rm k}})$ is also quite well known \citep{Kanan07, Ali14}.

\subsubsection{Cylindrically averaged Power Spectra}

Figure \ref{fig:ps2d_I+V_80sb} presents the cylindrically averaged power spectra
from the simulated visibilities using spherical harmonic ML inversion. We
calculated the power spectrum for two cases. The top panel of figure
\ref{fig:ps2d_I+V_80sb} shows the power spectrum corresponding to diffuse
foregrounds, 21-cm signal and noise (Stokes I), whereas the bottom panel
displays the noise-only power spectrum (Stokes V). We calculate the power
spectrum for a 12.5 MHz band around 150 MHz. We find the smooth diffuse
foreground in the Stokes I power spectrum mostly dominates at low
$k_{\parallel}$, where most of the foreground power is bound within
$k_{\parallel} \le 0.05 \rm{h cMpc^{-1}}$. We find the power drops by two to
three orders of magnitudes in high $k_{\parallel}$ regions, where the EoR plus
noise signal is expected to dominate. On the other hand, the Stokes V power
spectrum is more uniform and increases at higher $k_{\perp}$ scales where the
baseline density of LOFAR drops with respect to the central core region.

The mode-mixing introduced by the instrument chromaticity are usually confined
to a wedge-like structure in k space~\citep{Datta10,Morales12}. This wedge line
is defined by:

\be
k_{\parallel} = \left[\mathrm{sin}(\theta_{\mathrm{field}}) \frac{H_{0}D_{M}(z)E
(z)}{c(1+z)}\right]k_{\perp},
\e

where $\theta_{\mathrm{field}}$ is the angular radius of the FoV. In the power
spectra obtained from the same simulated visibilities but using the more
traditional method of gridding the visibilities in uv-space, a wedge like
structure is clearly visible (middle panel of Figure \ref{fig:ps2d_I+V_80sb}),
and is well known to be due to the frequency dependence of the
PSF~\citep{Vedantham12,Hazelton13}. Because our method consists of doing a
ML fit to non-gridded visibility data sets at each frequency, we effectively
obtain a PSF-deconvolved estimates of the sky spherical harmonics coefficients.
The mode-mixing due to the PSF frequency dependence is then considerably
reduced, demonstrated by the absence of a wedge in our ML power spectra
estimates (top panel of Figure \ref{fig:ps2d_I+V_80sb}).


\subsubsection{Spherically averaged Power Spectra}

\begin{figure}
    \includegraphics{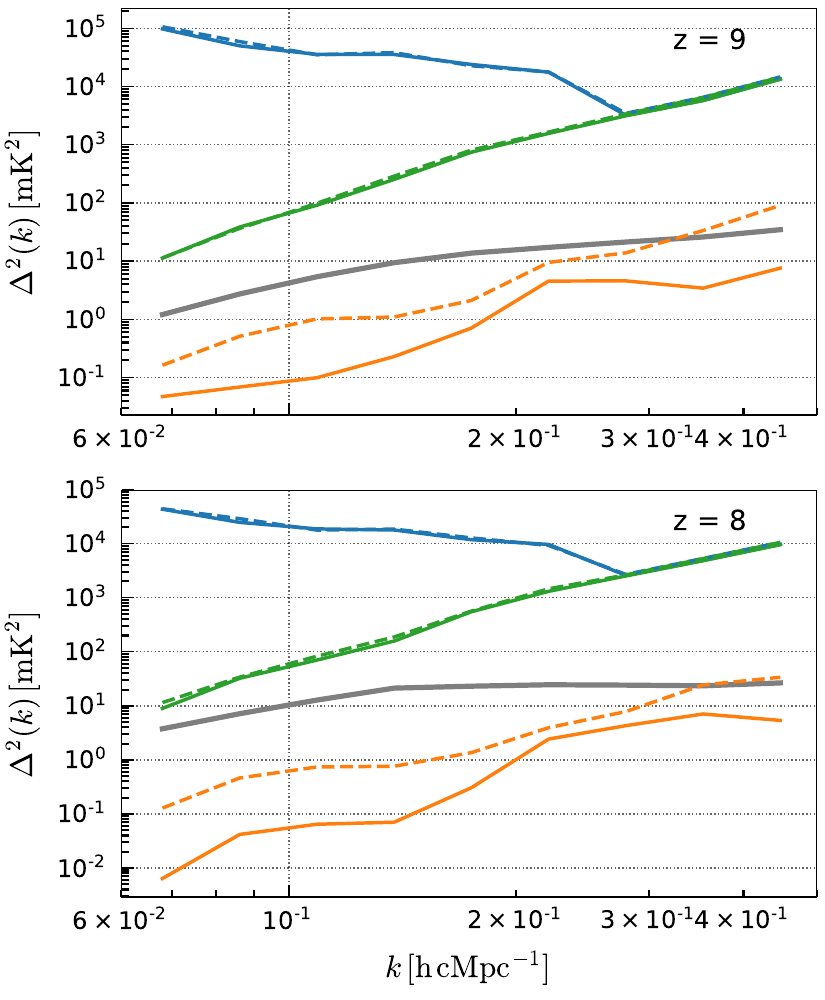}
    \caption{\label{fig:ps1d_sph_cart} Spherically averaged power spectra of Stokes
    I (blue) and Stokes V (green). The power spectra estimated using the spherical
    harmonics ML inversion (solid line) is compared to the power spectra estimated
    using the Cartesian ML inversion (dashed line). The difference between the input
    power spectra and estimated power spectra (orange) is lowered using the
    spherical harmonics ML inversion, and is well below the 21 cm signal power
    spectra (gray).}
\end{figure}

Next, we averaged the power spectrum in spherical shells and computed the
spherically averaged dimensionless power spectrum, $\Delta^2({\bf {\rm k}})={\rm
k^3P}({\bf {\rm k}})/2\pi^2$. In Fig. \ref{fig:ps1d_sph_cart} we present the
power spectrum in units of $\mathrm{mK}^2$ corresponding to Stokes I and Stokes
V visibilities. We observe that the spherically averaged power spectrum for the
Stokes I signal is mostly flat in the k range sampled by our simulation, whereas
the Stokes V or the noise power spectrum rises steeply from the low k to high k
values. It is also noteworthy that both the Stokes I and Stokes V spherically
averaged power spectra are at-least order of magnitude higher compared to the
21cm signal and hence the signal to noise (S/N ratio) is always low in our
simulation. 
In Fig. \ref{fig:ps1d_sph_cart} we compare our power spectrum estimates using
both the spherical harmonic ML inversion (solid line) and Cartesian ML inversion
techniques (dashed line) as introduced in \cite{Ghosh15} which we briefly
expose here. Ideally, the simulated visibility records a single mode of the
Fourier transform of the specific intensity distribution $I_{\nu}(\ell,m)$
corresponding to the simulated sky. Representing the celestial sphere by a unit
sphere, the component $n$ can be expressed in terms of $(l,m)$ by
$n(l,m)=\sqrt{1-l^2-m^2}$. Then the measured visibility for monochromatic,
unpolarized signal is,
\begin{eqnarray}
\begin{split}
\mathcal{V}_{\nu}(u,v,w) &= \int \frac{I_{\nu}(l,m)}{\sqrt{1-l^2-m^2}} \\ &\qquad \times 
e^{-2\pi i (ul+vm+w(\sqrt{1-l^2-m^2}-1))} \mathrm{d}l\mathrm{d}m.
\end{split}
\label{eq:Visibilityml}
\end{eqnarray}
We recall that Eqn. \ref{eq:Visibilityml} can be re-written in vector form
(similar to Eqn. \ref{eq:VisibMat}), where we used the Fourier kernel, $e^{-2\pi
i (ul+vm+w(\sqrt{1-l^2-m^2}-1))}$, is the response matrix $\responseSet$ and
$\frac{I_{\nu}(l,m)}{\sqrt{1-l^2-m^2}} \times d\Omega_p$ is the model sky
parameters that we want to directly infer from the visibility data. It is
interesting to point out that without any loss of generality (assuming that all
stations have identical primary beams) we can incorporate the Primary Beam (PB)
pattern in the response matrix \citep{Ghosh15} and  close to the phase center
the inverted specific intensity distribution closely follows to that where no
primary beam pattern is introduced. Hence, we decided not to introduce any PB in
the visibilities corresponding to the Cartesian ML solutions.

We notice that both Cartesian and Spherical-Harmonics ML inversion estimates are
quite close to each other across the whole k range sampled by our simulation.
This is also expected as we have restricted ourselves to a $\sim 4 \degree$
window (which corresponds to the full width half maxima
($\theta_{\mathrm{fwhm}}$) LOFAR-HBA stations) from the phase center where the
effects due to sky curvature is minimal. We also compared the difference between
the input and the reconstructed power spectrum from the spherical harmonic and
Cartesian ML inversion techniques. We note that the Cartesian ML approach
assumes a finite field for the sky model and hence does not account for
structure outside the FoV , hence leaves residual side-lobe noise. This is more
apparent for large angular scales where the Cartesian ML inversion error
increases compared to the SpH analysis, being the more appropriate choice for
large scales. We find that the difference between the input and the spherical
harmonic estimates (solid orange) is lower than difference between the input and
the Cartesian ML estimates (dashed orange line). For both redshifts z=8 and z=9,
the error due to the reconstruction lies well below the fiducial 21-cm signal,
which is shown in gray in Figure \ref{fig:ps1d_sph_cart}. This is an important
point as it shows that the error introduce by our SpH estimator will not affect
the statistical detection of the EoR signal across the whole k range probed by
the current simulation. Also, the SpH method could be further improved by
increasing the number of ($\ell, m$) modes beyond that currently set by our
sampling rule.

\section{Foreground removal algorithms} 

\begin{figure*}
    \includegraphics{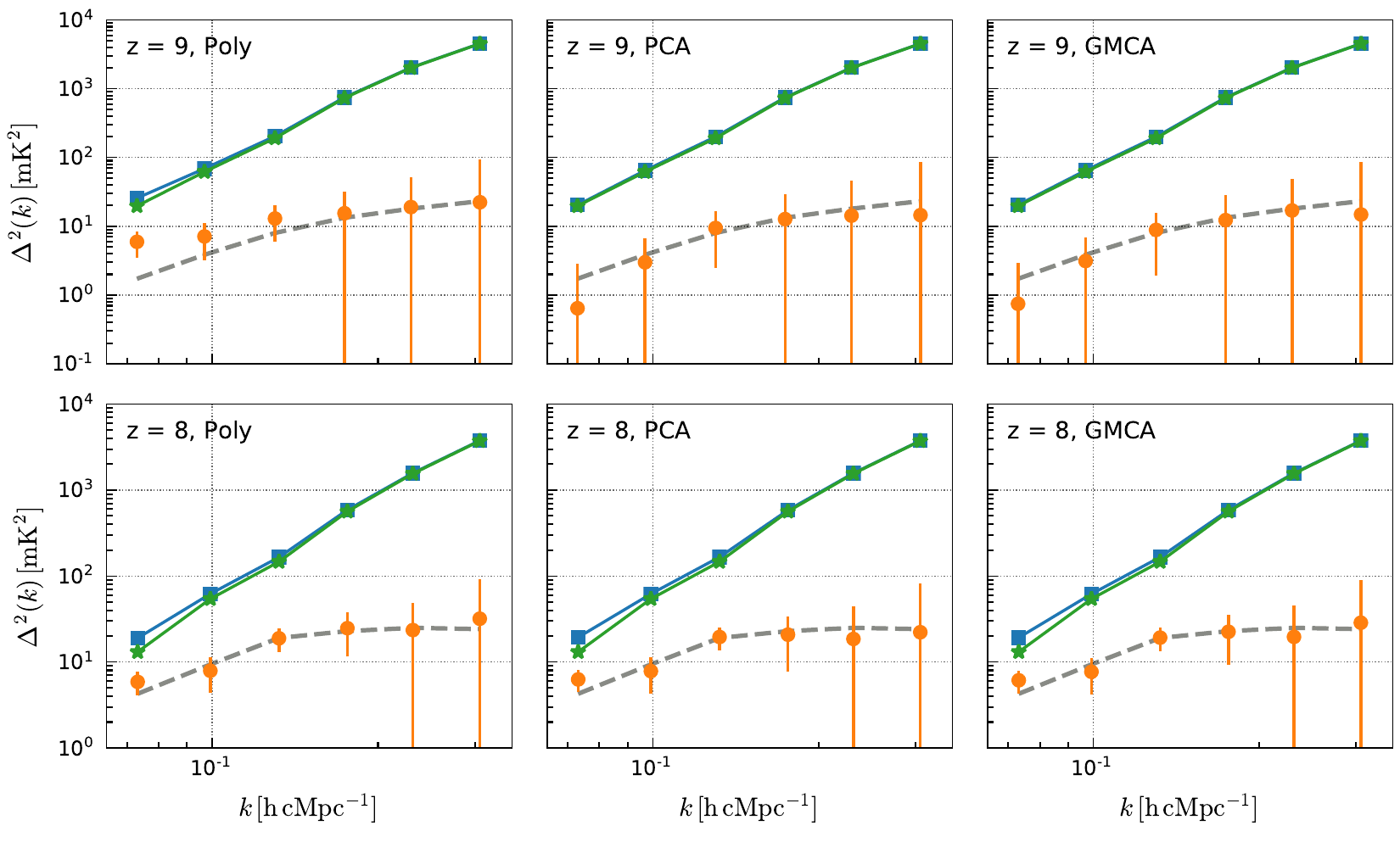}
    \caption{\label{fig:ps1d_eor2+3} Detection of the 21-cm
    signal using the spherically averaged power spectra. Subtracting the noise
    bias (green) from Stokes I residual power spectra after the foreground
    removal step (blue), we can recover the power spectra of the 21-cm line
    signal (orange) which is compared to the input 21-cm signal (gray dashed
    line). The error bar corresponds to 2-$\sigma$ sampling variance.}
\end{figure*}

Though the astrophysical foregrounds are expected to be approximately three to
four orders of magnitudes larger than the cosmological 21-cm HI signal, the two
signals have a markedly different frequency structure. The HI signal is expected
to be uncorrelated on frequency scales on the order of MHz, where as the
foregrounds are expected to be smooth in frequency. In this paper we have
implemented different foreground modeling techniques where for each ($\ell$, m)
mode we model the foreground as a smooth component in frequency. In the
following subsections, we briefly describe our foreground removal techniques.

\subsection{Polynomial Fitting} 

Probably, the most intuitively simplest method for
foreground removal is to choose an ad-hoc basis of smooth functions such as
polynomial fitting in frequency or log-frequency that we think can describe the
foregrounds \citep{McQuinn06, Morales06, Jelic08, Bowman06, Liu09}. Here, we use
polynomial fitting to fit each individual ($\ell$, m) mode along the line of sight direction. When fitting in log space, we offset the data to avoid negative
values. In log-log space our polynomial model is as follows:
\be
\mathrm{log}(b_{\ell m}(\nu)) = \sum_{k=0}^{N_{\mathrm{fg}}} \alpha_{\ell m}^{k}
[\mathrm{log}
(\nu)]^{k},
\label{polymodel}
\e
where, $N_{\mathrm{fg}}$ is the order of the polynomial and $\alpha_{\ell m}^{k}$ are the
coefficients of the polynomials.

One needs to carefully choose the order of the polynomial to avoid over or under-fitting  the foreground which could negatively affect the 21-cm EoR signal. Although, the  polynomial order can be chosen in a Bayesian manner where we can choose the particular polynomial model which has the highest evidence based on the data we have. In this paper,
we choose to fit 2nd order polynomial for each ($\ell, m$) modes whereas with both 
lower and higher order fits we find a worse result.

\subsection{Principal Component Analysis (PCA)}

Principal component analysis (PCA) utilizes the main properties of the
foregrounds such as their large amplitude and smooth frequency coherence to find
the largest foreground components and an optimal set of basis functions at the
same time \citep{Harker09, Masui13, Switzer13, Alonso15}. As the foreground are
highly correlated in frequency the frequency- frequency co-variance matrix of
the continuum foregrounds will have a particular eigen system where most of the
information can be sufficiently described by a small set of very large
eigenvalues, the other ones being negligibly small. Thus, we can attempt to
subtract the foregrounds by eliminating from the recovered ($\ell$, m) modes the
components corresponding  to the eigen-vectors of the frequency co-variance
matrix with the largest associated eigenvalues. In this paper, we choose to
remove 2 PCA components for each ($\ell$, m) mode which captured most of the variance of the foreground modes. This number will essentially depend on the frequency structure of the foregrounds and the different instrumental effects.

\subsection{Generalized Morphological Component Analysis (GMCA)}

GMCA is a blind source separation technique (BSS) \citep{Bobin07} which assumes
that a wavelet basis exists in which the smooth continuum foregrounds can be
sparsely represented with a few basis coefficients and thus can be separated
from EoR signal and noise \citep{Chapman13}. It is labeled as a non-parametric
method due to the lack of parameterized model for foregrounds which are largely
unknown at the low frequencies of interest. It uses the data to decide on the
foreground model. We note that GMCA is able to clean the foregrounds based on
both in spatial and frequency direction information contained within the
foreground signal compared to the cosmological signal and instrumental noise.
The result leads to a very different basis coefficients for the foregrounds and
the residual signal which is a combination of the method and the instrumental
noise.

\subsection{Application to the simulation}

In our simulation setup, we tried only to remove a minimal number of foreground
degrees of freedom (for example in case of the PCA method we look for the first two modes
corresponding to the highest variance) and thereby minimizing the risk of
subtracting the 21-cm EoR signal. We also assume that we know the noise variance
across the different k scales from Stokes V so that we can subtract the noise
power spectrum from the data and compare the residuals with the input 21-cm EoR
power spectrum. In Figure \ref{fig:ps1d_eor2+3} we show the recovered Stokes I
power spectrum from the SpH ML method, the Stokes V (noise power spectrum), the
residual (Stokes I - Stokes V) and the power spectrum corresponding to the 21-cm
EoR signal. We find, all the foreground removal methods recover the input power
spectrum quite well for $k \le 0.1 \rm{h\, cMpc^{-1}}$.

\begin{figure}
    \includegraphics{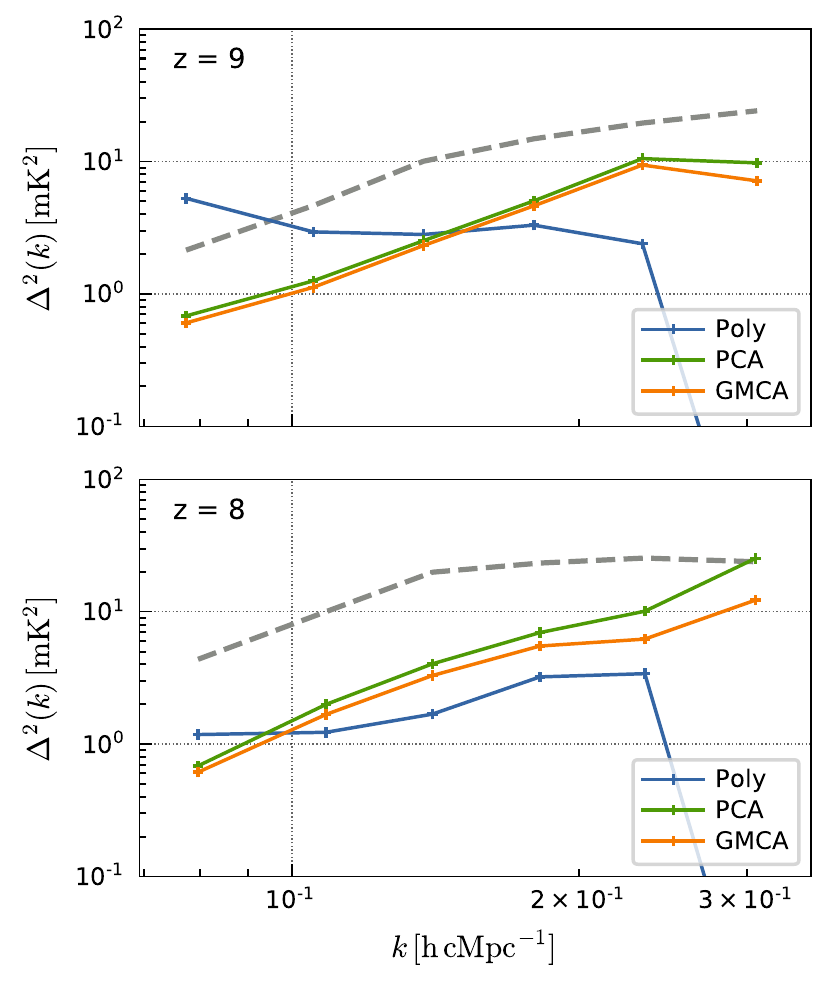}
    \caption{\label{fig:ps1d_fg_err} Spherically averaged
    power spectra of the foregrounds fitting error for the three different
    techniques used: the log-polynomial method (blue), the PCA method (green)
    and the GMCA method (red). The foregrounds fitting error is found to be
    below the 21-cm input signal (gray dashed line) in most situation except
    when using log-polynomial method at $z = 9$.}
\end{figure}

We observe that we can recover the input foreground model fairly well using all
three methods and that the differences between the input and recovered
foreground model are well below the fiducial 21-cm EoR signal at redshift z=8,
whereas we notice that at redshift z=9 the error is comparable to the EoR signal
below $k \le 0.1 \rm{h\, cMpc^{-1}}$, but lies well below the EoR signal for
higher $k$ ranges (Fig.~\ref{fig:ps1d_fg_err}). We also note all the three
foreground-removal approaches show similar extent of errors in reconstructing
the input foreground models, although at lower $k$ scales ($k \le 0.1 \rm{h\,
cMpc^{-1}}$) PCA and GMCA seem to work better than the polynomial fitting
method. We note this is mainly due to the differences in the three FG removal
methods and our SpH ML inversion method works perfectly well in reconstructing
the input signal across all the $k$ values probed by our current simulations
(Fig.~\ref{fig:ps1d_sph_cart}).

\section{Conclusions}

In this paper we have introduced a maximum likelihood spherical wave function
harmonic decomposition of the complex visibilities. The method can produce wide-
field images which will be a key component for next generation interferometers
with large field of view and new wide-field imaging challenges (ionosphere, beam
modeling etc.). The method has been formulated in a full sky setting including
the primary beam and its side-lobes, allowing us to model large parts of the sky
(up to a chosen $\theta_{\mathrm{max}}$) by considerably reducing the far side-
lobe noise which is an addition noise component due to the un-modeled structure
in the sky. We have shown, based on a spherical wave-function ML fit in the
visibility domain, that it is possible to deconvolve the chromatic ``wedge"
(caused by frequency-dependent side-lobes) in the ($\rm{k}_{\perp},
\rm{k}_{\parallel}$) power spectrum space, thus leaving us with a relatively
wider `clean' window to isolate the faint 21-cm EoR signal compared to the order
of magnitude strong foregrounds. This is particularly important when aiming to
achieve the expected level of sensitivity of future instruments such as the SKA,
for which it is expected that the far-out side-lobes of the station beam will
have a substantial impact on high dynamic range image performance
\citep{Cornwell92, McEwen08, Carozzi09, Carozzi15}.

We have shown that the coefficient of the visibility distribution in  spherical
coordinates is linearly related with the sky brightness distribution over a
celestial sphere. Hence, by decomposing the visibilities in spherical wave
functions, one provides a reconstruction of  the sky brightness distribution
without any extra computational cost. To reduce the computational load, we have
introduced a sampling scheme which speeds up the inversion considerably. In a
LOFAR-HBA full-sky simulation, including the 21-cm EoR signal, diffuse
foregrounds and a Gaussian random noise with rms of 0.04 Jy roughly
corresponding to 100 nights of 12 hr LOFAR integration time, we find that we can
recover the input power spectrum quite well across the whole $k$ range $0.07 -
0.3 \, \rm{h\, cMpc^{-1}}$. The foreground cleaning techniques implemented in
our current scheme works reasonably well and we notice that we can recover the
input EoR power spectrum assuming the noise power spectrum is known accurately.

Finally, we note that the simulated foregrounds and instrument model used in
this paper is not complete and does not include other foregrounds contaminants
such as the instrumental polarization leakage, the frequency dependence of the
individual LOFAR HBA station's primary beam and the phase errors caused by the
ionosphere or imperfect calibration. To tackle this problem, we are currently
working on a new foreground removal algorithm able to model multiple arbitrarily
non-smooth foreground contaminants, along with estimating their statistical error,
considerably improving the foregrounds model at the lowest $k$ where the 21-cm EoR 
signal also peaks, and delivering the full potential of the instrument.

The code implementing the algorithm described in this paper is freely available
at \url{https://gitlab.com/flomertens/sph_img}.

\section*{Acknowledgements}
AG would like to acknowledge Postdoctoral Fellowship from the South African Square Kilometre Array, South Africa (SKA-SA) for financial support. FMS and LVEK acknowledge support from a SKA-NL Roadmap grant from the Dutch ministry of OCW.




\bibliographystyle{mnras}
\bibliography{references} 

\bsp	
\label{lastpage}
\end{document}